\documentclass[aps,prl,floatfix,twocolumn]{revtex4}

\usepackage{amssymb}
\usepackage{amsmath}
\usepackage{amsfonts}
\usepackage{graphicx}
\usepackage{color}
\usepackage{xspace}
\usepackage{ulem}

\def\gsim{\raise0.3ex\hbox{$\;>$\kern-0.75em\raise-1.1ex\hbox{$\sim\;$}}}
\def\lsim{\raise0.3ex\hbox{$\;<$\kern-0.75em\raise-1.1ex\hbox{$\sim\;$}}}

\def \znbb {0\nu\beta\beta}

\newcommand{\AddrAHEP}{
  {\it AHEP Group, Instituto de F\'{\i}sica Corpuscular --
    C.S.I.C./Universitat de Val{\`e}ncia \\
    Edificio de Institutos de Paterna, Apartado 22085,
  E--46071 Val{\`e}ncia, Spain}}

\newcommand{\AddrDO}{%
Fakult\"at f\"ur Physik, Technische Universit\"at Dortmund, 
D-44221, Dortmund, Germany}

\newcommand{\AddrUFSM}{
Universidad T\'ecnica Federico Santa Mar\'\i a, \\ 
Centro-Cient\'\i fico-Tecnol\'{o}gico de Valpara\'\i so, \\ 
Casilla 110-V, Valpara\'\i so,  Chile}

\def\gsim{\raise0.3ex\hbox{$\;>$\kern-0.75em\raise-1.1ex\hbox{$\sim\;$}}}
\def\lsim{\raise0.3ex\hbox{$\;<$\kern-0.75em\raise-1.1ex\hbox{$\sim\;$}}}

\begin{document}

\preprint{IFIC/13-09}  

\title{Neutrinoless double beta decay and lepton number violation at the LHC}

\author{J.C. Helo} \email{juan.heloherrera@gmail.com}\affiliation{\AddrUFSM}
\author{M. Hirsch} \email{mahirsch@ific.uv.es}\affiliation{\AddrAHEP}
\author{H. P\"as}\email{heinrich.paes@uni-dortmund.de}\affiliation{\AddrDO}
\author{S.G. Kovalenko}\email{sergey.kovalenko@usm.cl}\affiliation{\AddrUFSM}

\keywords{double beta decay; neutrino masses and mixing; LHC}

\begin{abstract}
We compare the discovery potential of the LHC for lepton number
violating (LNV) signals with the sensitivity of current and future
double beta decay experiments, assuming $\znbb$ decay is dominated by
heavy particle exchange. We consider charged scalar, leptoquark and
diquark mechanisms of $\znbb$ decay, covering the $\znbb$ decay
operators with both, the smallest and largest, possible rates.  We
demonstrate, if $\znbb$ decay were found with a half-life below
$10^{26}-10^{27}$ ys a positive signal should show up at the LHC,
except for some particular cases of the leptoquark mechanism, and vice
versa, if the LHC does not find any hints for LNV, a ``short-range''
explanation for a finite $\znbb$ decay half-life will be ruled out in
most cases.  We argue, if a positive LNV signal were found at the LHC,
it is possible to identify the dominant contribution to $\znbb$.  Two
different kinds of observables which could provide such ``model
discriminating'' power are discussed: Different invariant mass peaks
and the charge asymmetry.
\end{abstract}

\maketitle

\section{Introduction} 
Neutrinoless double beta ($\znbb$) decay can be
induced by the exchange of light particles such as Majorana neutrinos
(long-range mechanism) or by heavy particles (short-range
mechanism). The classical example of the latter is right-handed
W boson and heavy neutrino exchange, which occurs in left-right
symmetric extensions of the standard model \cite{Riazuddin:1981hz}. A
number of other examples of such ``short-range'' contributions to
$\znbb$ decay have been discussed in the literature (for a review see
\cite{Deppisch:2012nb}), but only very recently the general
decomposition for the $\znbb$ decay operator, $ {\cal
O}^{0\nu\beta\beta}_{d=9} =\frac{1}{M^5} {\bar u}{\bar u}dd{\bar e} {\bar e}$
(where $M$ has dimension of mass) in terms of renormalizable vertices
connected by heavy virtual particles, has been given in
\cite{Bonnet:2012kh}.
Interestingly, several new diagrams for $\znbb$
decay were identified in this work, many of which require
quite exotic particles such as diquarks or colored fermions with
fractional charges. Common to all of the short-range contributions is
that masses of order $M \sim (1-5)$ TeV are needed for a $\znbb$ decay
half-life in the range of $T_{1/2}^{0\nu\beta\beta} \sim
(10^{25}-10^{27})$ ys in either $^{76}$Ge or $^{136}$Xe.

The very same diagrams which lead to $\znbb$ decay can lead to LHC
signals with like-sign dileptons plus (at least) two jets in the final
state \cite{Keung:1983uu,Allanach:2009iv}.  The ATLAS
\cite{ATLAS:2012ak} and the CMS \cite{CMS:PAS-EXO-12-017}
collaborations have published searches for such events. (Disregarding
flavor, we will refer to these events as ``$eejj$-like''.) ATLAS and
CMS then interpret their null result 
%
in terms of the left-right symmetric model 
as lower limits on the mass of the right-handed gauge boson 
$m_{W_R} \gsim (2.5-2.9)$ TeV \cite{CMS:PAS-EXO-12-017}, depending on
the mass of the right-handed neutrino $m_N$ 
and assuming the coupling 
$g_R$ of $W_R$ to fermions is equal to
the standard model coupling $g_L$. Note that \cite{ATLAS:2012ak}
provides separate limits for opposite-sign (OS) and same-sign (SS)
dileptons, while CMS provides only combined limits. While, of course,
the search for OS and SS is experimentally very similar, obviously
only SS-type events are a proof for the discovery of lepton number
violation (LNV).

In this paper we discuss how searches for LNV signals at the LHC
compare with the sensitivities of current and future $\znbb$ decay
experiments. We do not limit ourselves to a particular model of LNV,
but use a general decomposition of ${\cal O}^{0\nu\beta\beta}_{d=9}$.
%
We show that 
a nonobservation of $eejj$-like events in the future LHC run at
$\sqrt{s}=14$ TeV would rule out all but the pure leptoquark
decompositions as the dominant contribution to ${\cal O}^{0\nu\beta\beta}_{d=9}$. 
Perhaps even more interesting is that, if
a positive LNV signal were found at the LHC, it is possible to
identify the dominant contribution to $\znbb$. We discuss two types of
observables which could provide such ``model discriminating'' power:
The construction of different invariant mass combinations and the
ratio of positronlike to electronlike events. We will call the
latter the ``charge asymmetry''.

\section{Short-range double beta decay} 
\begin{figure}[htb]
\centering
\begin{tabular}{cc}
\includegraphics[width=1.0\linewidth]{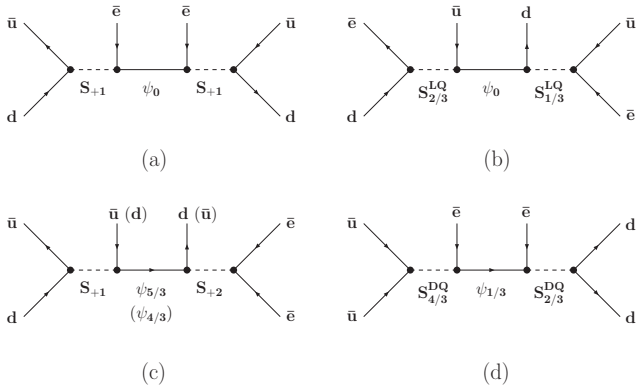}
\end{tabular}
\vskip-3mm
\caption{
%
Example diagrams of Topology I for short-range double beta decay, see text: 
                     (a) charged scalar $S_{+1}$ -- neutral fermion $\psi_{0}$ -- charged scalar $S_{+1}$;
                     (b) leptoquark $S^{LQ}_{2/3}$ -- neutral fermion $\psi_{0}$ -- leptoquark $S^{LQ}_{1/3}$;
                     (c) charged scalar $S_{1}$ -- charged colorful fermion $\psi_{q}$ -- charged scalar $S_{+2}$; 
                     (d) diquark $S^{DQ}_{4/3}$ --  charged colorful fermion $\psi_{1/3}$ -- diquark $S^{DQ}_{2/3}$
}
\label{fig:Diags}  
\vskip-3mm
\end{figure}

Double beta decay can be induced by exchange of scalars or vectors with or without an
accompanying virtual fermion.  Since the results for scalars and vectors
are qualitatively very similar, we will concentrate on the scalar case
only. Also, there are two basic topologies of the diagrams
contributing to $0\nu\beta\beta$ decay. Here, we limit ourselves to
the discussion of Topology I (scalar-fermion-scalar exchange) shown in
fig. (\ref{fig:Diags}). The results for Topology II (triple-scalar
exchange coupled to the trilinear vertex) will be presented elsewhere
\cite{Helo:2013xx}.
There are in total 18 different possible decompositions (``diagrams'')
in Topology I that can contribute to the $\znbb$ decay operator,
${\cal O}^{0\nu\beta\beta}_{d=9}$.  Depending on the color and
$SU(2)_L$ transformation properties of the internal fermion and
bosons, each decomposition can be realized in several ways, see
below. Twelve decompositions involve fractionally charged (and colored)
fermions, and were discussed for the first time in
\cite{Bonnet:2012kh}. The remaining six possibilities have all been
discussed before in the literature in the context of concrete models,
most notably the mass mechanism or R-parity violating supersymmetry
\cite{Mohapatra:1986su,Hirsch:1995zi}.

In Fig.1 we show some example diagrams for $\znbb$
decay. These examples were chosen such that all six different bosons
($S_{+1}$, $S_{+2}$, $S^{LQ}_{2/3}$, $S^{LQ}_{1/3}$, $S^{DQ}_{4/3}$
and $S^{DQ}_{2/3}$) and all four types of fermions ($\psi_0$,
$\psi_{1/3}$, $\psi_{4/3}$ and $\psi_{5/3}$), which can contribute to
$\znbb$ decay, appear at least once each in the figure. We use the
notation $S^{x}_q$ (and $\psi_q$) to denote a scalar (fermion) with
electric charge $q$ and the superscript $x=DQ$ or $x=LQ$ to
differentiate between diquarks and leptoquarks.  Note that replacing
$S_{+1}$ in Fig.1(a) on both sides by a standard model W boson and
interpreting $\psi_0$ as a neutrino would lead to the classical mass
mechanism, and diagrams of the type shown in Fig.1(b) appear, for example,
in R-parity violating supersymmetry, while Figs.1(c) and 1(d) are examples of
the new type of diagrams.

Integrating out the heavy particles in all cases leads to  the
${\cal O}^{0\nu\beta\beta}_{d=9}$ operator and to 
an amplitude of $\znbb$ decay
\begin{equation}\label{eq:dim9}
{\cal A}^{0\nu\beta\beta}
\propto 
\frac{g_1g_2g_3g_4}{m_{S_i}^2m_{\psi_q}m_{S_j}^2} 
\equiv \frac{g_{eff}^4}{M_{eff}^5}.
\end{equation}
We denote the couplings as $g_1$-$g_4$, arbitrarily read from
left to right in the diagrams. All of these are in principle different
free parameters in general. $m_{S_i}$, $m_{S_j}$ and $m_{\psi_q}$ are
the masses of the bosons and the fermion, respectively, not necessarily 
$m_{S_i}=m_{S_j}$. 

While ${\cal A}^{0\nu\beta\beta}$ depends on the particle
physics parameters always in the form of Eq.(\ref{eq:dim9}), the
nuclear matrix elements (and thus the half-life) depend on the
chirality of the ``outer'' fermions as well as the color
representation
of the ``inner'' particles in the
diagram. The recent paper \cite{Bonnet:2012kh} provides a complete
list of operators (for short-range $\znbb$ decay with scalars), which,
once numerical values for the nuclear matrix elements are taken from,
for example \cite{Deppisch:2012nb}, can be directly converted into
calculated half-lives, using $T^{\znbb}_{1/2} = {\cal C}_i^2 ({\cal
M}_i^{Nucl})^2 G$, where $G$ stands for a phase space integral, ${\cal
M}_i^{Nucl}$ for the nuclear matrix element and ${\cal C}_i$ is the
coefficient of the dimension-9 operator made dimensionless by a numerical
factor of $2 m_P/G_F^2$.  Current limits from $^{76}$Ge
\cite{KlapdorKleingrothaus:2000sn} and $^{136}$Xe
\cite{Auger:2012ar,Gando:2012zm}, both of the order of
$T_{1/2}^{\znbb}\gsim 10^{25}$ ys, correspond to roughly $M_{eff}
\gsim (1.2-3.2) g_{eff}^{4/5}$ TeV, depending on decomposition.  These
limits depend on nuclear matrix elements, for which a typical uncertainty 
of a factor of 
$\sim$2
is usually quoted 
in the literature.

\section{Sensitivity: LHC versus $\znbb$ decay}
The number of $eejj$-like events at the LHC in general depends on a 
different combination of couplings and masses than the $\znbb$ decay 
amplitude. Here, for brevity, we will discuss only the ``symmetric 
decompositions'', see below. We stress, however, that we have checked 
that despite being a special case, the results presented below still 
cover both the most optimistic and the most pessimistic scenario 
for the LHC. A systematic case-by-case study covering all decompositions 
and cases in detail will be presented elsewhere \cite{Helo:2013xx}. 
In ``symmetric'' decompositions, such as Fig.1(a), couplings are pairwise equal, i.e. $g_1=g_4$ and 
$g_2=g_3$. Assuming that such an equality holds at least
approximately, in the small width approximation we can write the
number of events at the LHC as 
\begin{eqnarray}\label{eq:SigBr}
\#(eejj)/\cal{L} &= & \sigma(pp\to S_i)\times{\rm Br}(S_i\to eejj) \\ 
\nonumber
 = & F_{S_{i}}&
\left( \sqrt[4]{\frac{(M_{\text{eff}(S_{i})})^5}{m_{\psi_q}}}\right)  
g_{eff}^4 {\rm Br}^{eff}(S_i\to eejj).
\end{eqnarray}
It is assumed here that $m_{S_i}>m_{\psi_q}$, such that both particles
are on shell. Thus, our analysis is not valid for $m_{S_i}
<m_{\psi_q}$. Also, we have rewritten $F_{S_{i}}(m_{S_i})=\sigma(pp\to
S_i)/g_1^2$ and ${\rm Br}^{eff}(S_i\to eejj)={\rm Br}(S_i\to
eejj)/g_2^2$. While ${\rm Br}(S_i\to eejj)$ is fixed for a given
decomposition, once all couplings and masses are fixed, here we simply
treat ${\rm Br}^{eff}(S_i\to eejj)$ as a free number smaller than
one. Note, however, that in the limit where all couplings are equal
(and $m_{\psi_q}=0$) ${\rm Br}(S_{+1}\to e^+e^+jj)$= \mbox{${\rm
Br}(S_{+1}\to e^+e^-jj)$} $\simeq 1/8$ for $S_{+1}$.  In the LQ case,
the LQ is produced in association with a lepton, i.e. we calculate
$\sigma(pp\to S_i+e)\times {\rm Br}(S_i\to ejjj)$, see next section.
Single LQ production at the LHC is mainly through $G+q \to
S^{LQ}_q +e$, i.e. to the corresponding diagrams in
Fig. (\ref{fig:Diags}) one has to attach a gluon to the initial 
quark.  We have used CalcHEP \cite{Pukhov:2004ca} and MadGraph 5
\cite{Alwall:2011uj} to calculate the production cross sections for
$S_{+1}$, $S^{LQ}_q$, and $S^{DQ}_q$ at the LHC. We have compared our
results with the literature
\cite{Ferrari:2000sp,Belyaev:2005ew,Han:2010rf} and found quite good
agreement in all cases.

\begin{figure}[htb]
\centering
\begin{tabular}{cc}
\hskip7mm\includegraphics[width=0.82\linewidth]{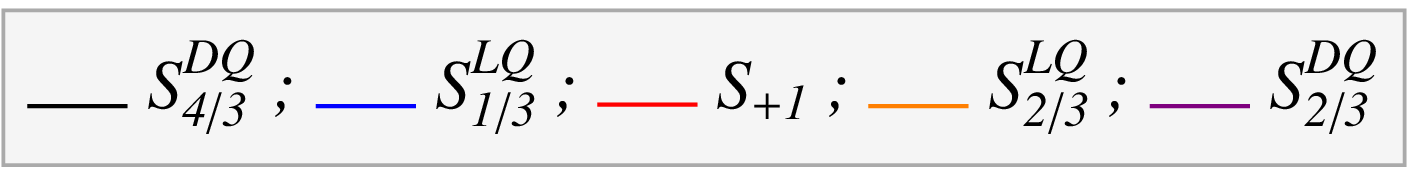}\\
\includegraphics[width=0.95\linewidth]{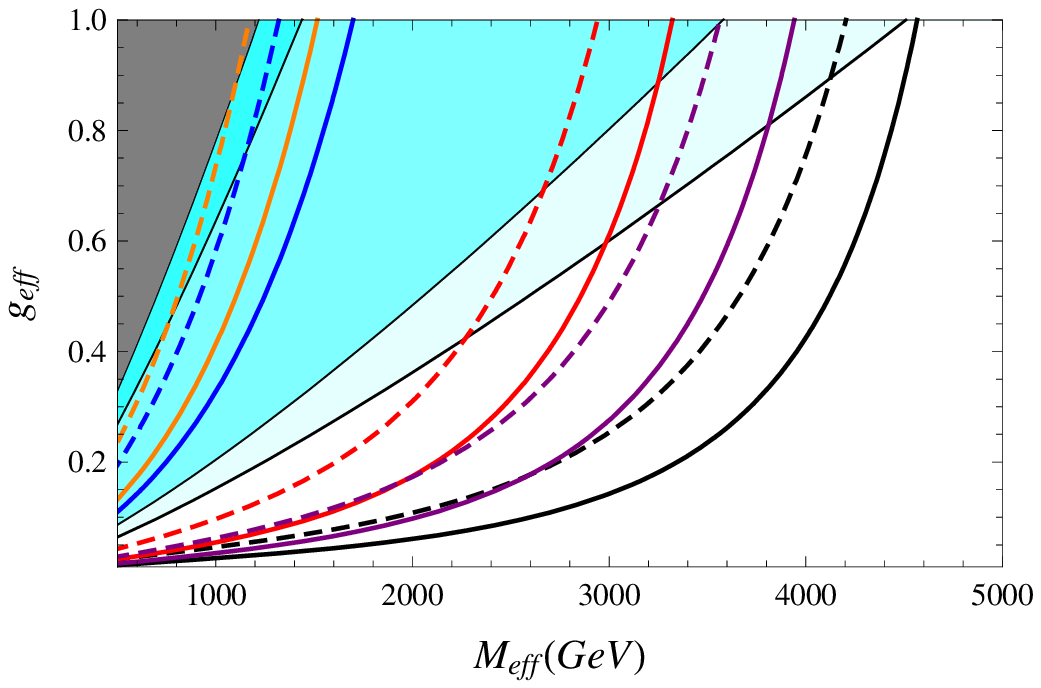} \\
\includegraphics[width=0.95\linewidth]{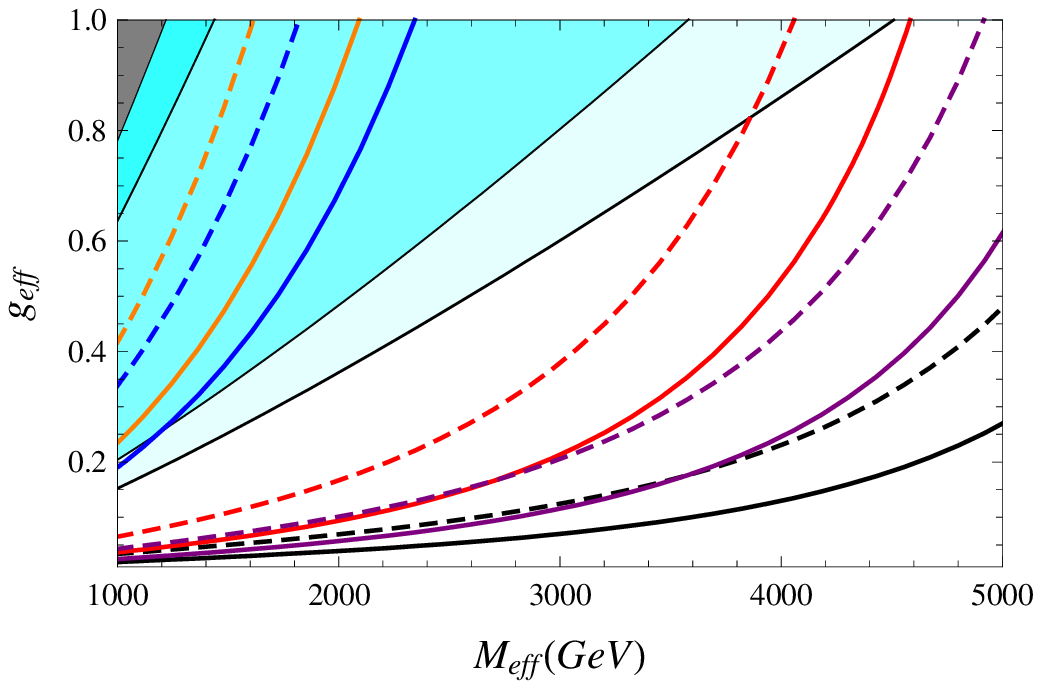}
\end{tabular}
\vskip-3mm
\caption{Sensitivity limits for the LHC for production of five different 
scalar bosons compared to current and future double beta decay experiments. 
(a) Top: $m_{\psi_q}=200$ GeV; (b) bottom: : $m_{\psi_q}=1$ TeV, 
see text.}
\label{fig:Lmts}  
\vskip-3mm
\end{figure}  

In Fig. (\ref{fig:Lmts}) we then plot the compared sensitivities of
$\znbb$ decay and the LHC for five different cases, using
Eq. (\ref{eq:SigBr}). The more complicated ``asymmetric'' case, with
more than one scalar mass and all couplings different, will be
presented elsewhere \cite{Helo:2013xx} but leads to the same
conclusions. 
For the LHC we show the expected sensitivity, assuming less
then three signal events in 300 fb$^{-1}$ of statistics.
Note, however, that our conclusions are not affected even if this assumed upper limit is as large as ten signal
events. We plot for two values of ${\rm Br}^{eff}(S\to eejj)$,
i.e $10^{-2}$ (dashed lines) and $10^{-1}$ (solid
lines) in the plane $g_{eff}$ versus $M_{eff}$, for two different
values of $m_{\psi}=200$ GeV (top) and $m_{\psi}=1$ TeV (bottom).
The different color codes of the lines correspond to the five different scalar bosons,
that can be singly produced at the LHC, namely $S_{+1}$,
$S^{LQ}_{2/3}$, $S^{LQ}_{1/3}$, $S^{DQ}_{4/3}$ and $S^{DQ}_{2/3}$.
Note that $S_{+1}$ is very similar to the well-known case of $W_R$
production in LR symmetry and $S_{2/3}^{LQ}$ and $S_{4/3}^{DQ}$
correspond to the most pessimistic and most optimistic scenario for
the LHC.

In addition, Fig. (\ref{fig:Lmts}) shows four different cases for
current and future limits from $\znbb$ decay. The dark gray area
is the currently excluded part of parameter space from nonobservation
of $^{136}$Xe decay with $T_{1/2}^{\znbb}\ge 1.6 \times 10^{25}$ ys
\cite{Auger:2012ar} assuming $\znbb$ decay is caused by the
operator with the smallest rate and thus corresponds to the most
pessimistic case for the sensitivity of $\znbb$ decay. The three blue
areas are (from left to right) the smallest rate, but for a limit of
$T_{1/2}^{\znbb}\ge 10^{26}$ ys, the largest rate with
$T_{1/2}^{\znbb}\ge 10^{26}$ ys, and the largest rate with
$T_{1/2}^{\znbb}\ge 10^{27}$ ys. The lightest area to the right
therefore corresponds to the most optimistic reach for $\znbb$ decay
in the foreseeable future.

As can be seen, with the exception of the LQ cases, the LHC at
$\sqrt{s}=14$ TeV will be more sensitive than $\znbb$ decay
experiments as probe for LNV. For the LQ case, the LHC covers 
the region explored by future $\znbb$ decay experiments only 
partially. Note that decompositions with one LQ
and one $S_{+1}$ or one $S^{DQ}_{q}$ will be similarly constrained by
LHC searches as the ``symmetric'' $S_{+1}$ or $S^{DQ}_{q}$ case.

\section{Distinguishing LNV models}
As long as LHC experiments provide only upper limits on LNV signals, 
conversion to model-dependent exclusion plots is, apart from different 
numerical factors, quite similar in all cases. However, if a positive 
signal is observed in the future, the LHC has the potential to differentiate 
between different LNV models. Probably the two most promising observables 
to do so are the different {\it invariant mass peaks} and  the {\it charge 
asymmetry}.

\subsection{A. Invariant mass peaks}
One can divide the discussion into (i) s-channel production 
of either $S_{+1}$ or $S^{DQ}_{q}$, (ii) associated $S^{LQ}_{q} + e$ 
production and (iii) pair production of a colored fermion $\psi_q$. 

(i) All decompositions with s-channel production of $S_{+1}$ or
$S^{DQ}_{q}$ lead to the final state $eejj$ with a mass peak at
$m_{eejj}^2=m_S^2$.  Consider first the simplest case, Fig.1(a). 
This case is equal to the well-known
LR-symmetric case: On-shell production of $S_{+1}$ decaying to
$\psi_0+e$, with $\psi_0\to ejj$ via off-shell $(S_{+1})^*$ leads to
two invariant mass peaks $m_{eejj}^2=m_{S_{+1}}^2$ and
$m_{e_2jj}^2=m_{\psi_0}^2$. On the other hand, $m_{ee}^2$ has a broad
distribution. Compare this to, for example, Fig.1(c). 
Here, still $m_{eejj}^2=m_{S_{+1}}^2$, but
$m_{eej_2}^2=m_{\psi_q}^2$ and since $m_{S_{+2}}<m_{\psi_q}$ is
possible, a third peak could show up at $m_{ee}^2=m_{S_{+2}}^2$. 

(ii) $S_{q}^{LQ}$ production is different, since it is produced in
association with a lepton, i.e. there are (at least) three jets in the
final state and $m_{eejjj}^2$ does {\em not} show a peak, 
while $m_{e_2jjj}^2=m_{S_{i}^{LQ}}^2$.  Again, different subsystems of
$e_2jjj$ have peaks, depending on mass ordering and exact
decomposition, but in the $S_{q}^{LQ}$ case there never is a peak in the
$m_{ee}^2$ distribution.  

(iii) As mentioned above, 12 decompositions have fractionally
charged fermions, all of which have to be colored. These fermions can
be pair produced through gluon-gluon fusion and the
final state of interest to us is $ee+4j$. The events of this type will
have a threshold of 
$m_{eejjjj}^2= 4 m_{\psi_q}^2$ 
and their subsystems either $eej$ {\em and} $jjj$ or $e_{1}j_{1}j_{2}$ {\em and} $e_{2}j_{3}j_{4}$
will show mass peaks at $m^{2}_{\psi_{q}}$.

Finally, there are decompositions with both diquarks and leptoquarks with
in general different masses, which could lead to  events with $2j$, $3j$ and $4j$ 
having mass peaks in different subsystems and
at different values.
%


The following comments might be in order. 
First, charged scalar and both diquarks leave the ``classical'' mass
peak in $eejj$. The experimental sensitivity to this mass peak and the
experimental resolution for it is known, and has been simulated by
both ATLAS  \cite{ATLAS:2012ak} and CMS \cite{CMS:PAS-EXO-12-017}. 
The different subsystems, that we discussed as ``model'' discriminators are then found
within the set of those events.  Thus, their energy spread is determined
by and smaller than the one of the full system $eejj$.  
The only mass peaks, which are indeed different, are the ones produced by leptoquarks,
where the events $eejjj$ are more spread out in energy. For obtaining
the exact number of events, which are needed for discovery, a MonteCarlo
simulation will be helpful and will be carried out elsewhere.
\begin{figure}[htb]
\centering
\begin{tabular}{cc}
\includegraphics[width=0.95\linewidth,height=0.85\linewidth]{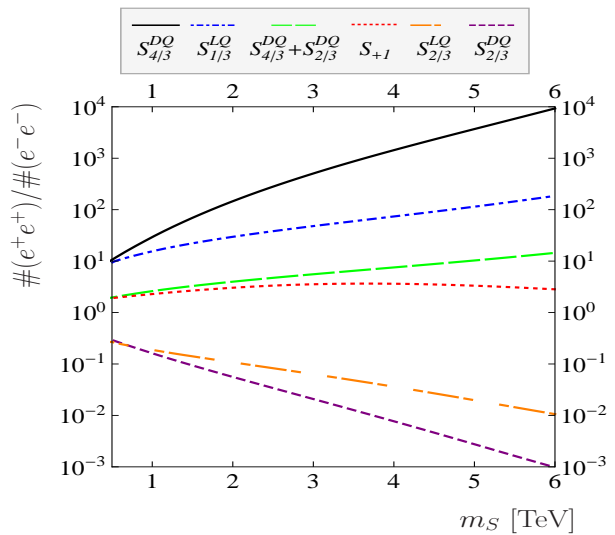}
\end{tabular}
\vskip-3mm
\caption{Charge asymmetry, i.e. ratio of positronlike to electronlike 
events as a function of the boson mass for different kinds of scalars.}
\label{fig:CASmplst}  
\end{figure}  

\subsection{B. Charge asymmetry}
Define the charge asymmetry as the ratio of the number of events 
\begin{equation}\label{eq:CA}
x_{CA} = \#(e^+e^+)/\#(e^-e^-).
\end{equation}
Consider first the simplest case of Fig.1(a).  With only one boson and two pairwise equal
couplings ($g_1=g_4$ and $g_2=g_3$) $x_{CA}$ is simply determined by
the ratio of production cross section for $\sigma(pp\to S_{+1})$
divided by $\sigma(pp\to S_{-1})$. Since the number of $u$ quarks is
different from the number of $d$ quarks in the proton, the two cross
sections are not the same and one expects to see more events with
$e^+e^+$ than $e^-e^-$. More generally, $x_{CA}$ is a function of 
all masses and couplings, but even in the limit where couplings 
are equal, different $\znbb$ decay decompositions have different 
values for $x_{CA}$.

Figure (\ref{fig:CASmplst}) shows the theoretically expected value of
$x_{CA}$ for six different cases in the equal coupling limit. These 
cases shown do not (always) directly correspond to a particular 
decomposition, but represent extreme limits, where the scalar 
identified by the color code gives the dominant contribution to $x_{CA}$. 
The case
$S_{2/3}^{DQ}+S_{4/3}^{DQ}$ assumes that both diquarks have the same
mass and couplings and thus is just the average of the two individual
cases. If both $S_{1/3}^{LQ}$ and $S_{2/3}^{LQ}$ contribute equally to
the $\znbb$ decay rate (case not shown), the two asymmetries average
to a number close to, but not equal to the case of $S_{+1}$.  Note
that the case for $S_{+1}$ is equal to the expectations in the
left-right symmetric model. In this cases, the charge asymmetry varies
only weakly from $x_{CA} \simeq (2-4)$ in the mass range
shown. However, as the figure demonstrates, in other cases much larger
(and much smaller) $x_{CA}$ are possible. Experimentalists should
therefore provide data sets separately for positively and negatively
charged leptons, in order to cover all LNV models.
Finally, note that for the LQ case we show curves up to 6 TeV, but
realistically (compare Fig. (\ref{fig:Lmts})) for LQs the LHC will not
have sufficient sensitivity to find any events for LQ masses larger
than roughly 2 TeV.

\section{Conclusions}
We have compared the sensitivity of the LHC experiments for LNV
signals with current and future sensitivities of $\znbb$
decay experiments. The LHC has large discovery potential for general LNV violating
models. In case a positive signal is found in the future at the LHC,
we have proposed two 
types of observables, which allow distinguishing
different LNV models: (i)
particular sets of
invariant mass peaks and (ii) the ratio of positron to electron same-sign dilepton events.

\medskip
\centerline{\bf Acknowledgements}

This work was supported by UNILHC PITN-GA-2009-237920
and by the Spanish MICINN Grants No, FPA2011-22975 and No. MULTIDARK
CSD2009-00064, by the Generalitat Valenciana (Prometeo/2009/091), by Fondecyt
Grants No. 11121557 and No. 1100582, and  CONICYT Projects
No. 791100017 and No. CONICYT/DFG-648.  H.P. was supported by DGF Grant No. PA 803/6-1.

\end{document}